\documentclass{iopjournal}
\usepackage[backend=biber,natbib,style=nature,doi=true]{biblatex}
\usepackage{amsmath}
\usepackage{amssymb}
\usepackage{siunitx}
\usepackage{tabularx}

\usepackage[version=4,arrows=pgf-filled,mathfontname=mathsf]{mhchem}
\usepackage{graphicx}
\usepackage{bm}
\usepackage{siunitx}
\usepackage{xspace}
\usepackage{textcomp}
\usepackage[gen]{eurosym}
\usepackage{adjustbox}

\begin{document}

\articletype{Letter}

\title{Verification and experimental validation of neutral atom beam source produced by L-PBF}

\author{Vineet Kumar$^1$\orcid{0000-0001-8668-3663},
Niklas V. Lausti$^1$\orcid{0000-0001-9906-6971},
Peter K\'{u}\v{s}$^1$\orcid{0000-0002-2246-4426},
Adam Jel\'{i}nek$^1$\orcid{0009-0004-8659-4539},
Ivan Hud\'{a}k$^{1,2}$\orcid{0009-0004-4162-6154},
David Moty\v{c}ka$^1$\orcid{0009-0005-9012-4721},
Petr Dohnal$^1$\orcid{0000-0003-0341-0382},
Radek Pla\v{s}il$^1$\orcid{0000-0001-8520-8983},
Ji\v{r}\'{i} Hajny\v{s}$^3$\orcid{0000-0002-9228-2521},
Michal Hejduk$^1$\orcid{0000-0002-4417-4817}
}

\affil{$^1$Charles University, Faculty of Mathematics and Physics, Dept.\ of Surface and Plasma Science, Prague 8, Czech Republic}

\affil{$^2$Institute of Photonics and Electronics CAS, v.v.i., Chaberská 1014/57, Prague 8, Czech Republic}

\affil{$^3$Faculty of Mechanical Engineering, VŠB - Technical University of Ostrava, Ostrava, Czech Republic}

\email{michal.hejduk@matfyz.cuni.cz}


\begin{abstract}

We report validation tests of a calcium atomic-beam source fabricated via Laser Powder Bed Fusion (L‑PBF). The surface quality and elemental composition of the printed component were quantitatively assessed, allowing us to establish reference parameters for reliable operation in an ultra-high-vacuum environment. Safe operating conditions of the atomic oven were determined through a combination of simulations and experimental measurements. The ability of the device to deliver an atomic beam to the main experimental region -- the electron/ion trap -- was verified using atomic fluorescence imaging. Fluorescence spectroscopy was further employed to characterize the beam divergence, yielding an emission-cone half-angle of approximately \qty{19}{\degree} for atoms near the beam axis. A current of atoms on the order of \qty{E8}{\per\second} was estimated in the electron-trapping region, which is more than sufficient for anticipated electron-trapping and ion-trapping experiments.

\end{abstract}


\section{Introduction}
\label{sec:Introduction}

When designing new experiments with laser-cooled atomic ions for applications in quantum chemistry\cite{miossec_design_2022}, quantum sensing\cite{gilmore2021quantum}, and quantum computing\cite{manovitz_trapped-ion_2022}, one critical early consideration is determining how to produce the ions. Neutral atoms can be released from solid targets through either laser ablation\cite{hendricks2007all} or thermal evaporation\cite{ross1995high}.

Laser ablation offers temporal control over ion production on the order of \qty{1}{\nano\second} (duration of laser pulses). However, this method requires direct optical access to the target and necessitates shifting the focus of the laser over the target to maintain the reproducibility of the atom generation\cite{hendricks2007all,duncan_invited_2012}. Additionally, it lacks precise control over the species of particles produced\cite{olmschenk_laser_2017}.

For thermal evaporation methods, the primary challenge lies in controlling continuous particle production. The solution centres on effective thermal management through rapid evaporation\cite{gao2021optically} of well-insulated source material\cite{verma2017compact,ballance2018short}, supplied in minimal quantities to prevent excessive production\cite{schwindt2016highly}.

Among the two most widely used heating techniques -- laser irradiation\cite{gao2021optically} and Joule heating\cite{nomura_direct_2023} -- the latter offers minimal technical complexity. Nevertheless, designing an atom source can be challenging when thermal radiation risks influencing experimental outcomes. For instance, in experiments involving trapped electrons generated by photoionization of atoms\cite{mikhailovskii_trapping_2025,kumar2025vacuum}, radiation from the oven and adjacent components may heat the trap electrodes. This is critical because the decoherence rate of a trapped-electron quantum harmonic oscillator due to Johnson–Nyquist noise scales with the square of the temperature\cite{brownnutt_ion-trap_2015}. A shift from room temperature to the operational temperature of the oven -- \qty{600}{\kelvin} for calcium -- can increase the decoherence rate by a factor of four.

Conversely, the oven aperture must be positioned close to the trap centre to ensure a high density of evaporated atoms, while simultaneously accommodating spatial constraints imposed by internal components and preventing unwanted deposition of atoms on them. A proper design can balance these two competing requirements -- proximity of the oven to the electron trap versus thermal isolation -- but achieving such a design using traditional fabrication methods often encounters geometrical and financial limitations. To overcome these challenges, we previously investigated whether Laser Powder Bed Fusion (L-PBF) could enable the production of a functional oven. While \citet{norrgard2018note} introduced a 3D-printed lithium dispenser for cold alkali atom experiments in magneto-optical traps, our earlier work demonstrated that larger structures produced by the L-PBF method are compatible with ultra-high vacuum and the requirements of electron trapping experiments\cite{kumar2025vacuum} .

Building on our earlier demonstration of the feasibility of using Laser Powder Bed Fusion (L-PBF) to fabricate a functional oven for neutral atom production, the present work advances this concept by addressing two critical aspects: (i) the design rationale for balancing proximity to the trap centre with thermal isolation, and (ii) a thorough verification of the manufactured component. This includes dimensional, qualitative, and compositional analysis, as well as experimental validation of its performance as a neutral atom beam source through direct fluorescence observation and beam alignment toward the trap centre. By providing these insights, we establish not only the practicality of additive manufacturing for complex vacuum-compatible components but also its reliability for precision atomic experiments.

\section{Construction of the oven assembly} \label{sec:Construction of the oven assembly}

Our experiment is designed to use calcium, which has the vapour pressure of $\sim \qty{E-3}{Pa}$ at the temperature of \qty{685}{K}. To achieve the atom flux high enough for experimental purposes, the oven has to be operated at even higher temperatures. The 316L grade stainless steel we have chosen as the main material can certainly stand such temperatures without changes in the physical properties and provides relatively low thermal conduction. A ceramic like aluminium nitride, also printable\cite{belmonte_heat_2021}, is an alternative to the stainless steel (SS) but is more brittle and provides elevated danger of outgassing. The employment of 316L SS, on the other hand, requires usage of electrical insulators made from the ceramics, but those can be purchased off the shelf.

Between the options of using an off-the-shelf calcium powder dispenser or a custom-made one, we chose the former (AlfaVakuo e.U., type A12), which comes with an indium seal. This seal can be melted by Joule heating inside the UHV apparatus, preventing excessive oxidation of the calcium powder during filling and thereby avoiding adverse impacts on calcium ion production. However, the dispenser part can be replaced with a custom-made part at any time\cite{norrgard2018note}.

Now that we have the constraints to the material and parts, we are ready to immerse into the particularities of the design.

\subsection{Designing and 3D-printing of oven assembly}
\label{sec:Designing and 3D-printing of oven assembly}

\begin{figure*}
    \centering
    \includegraphics[width=0.9\textwidth]{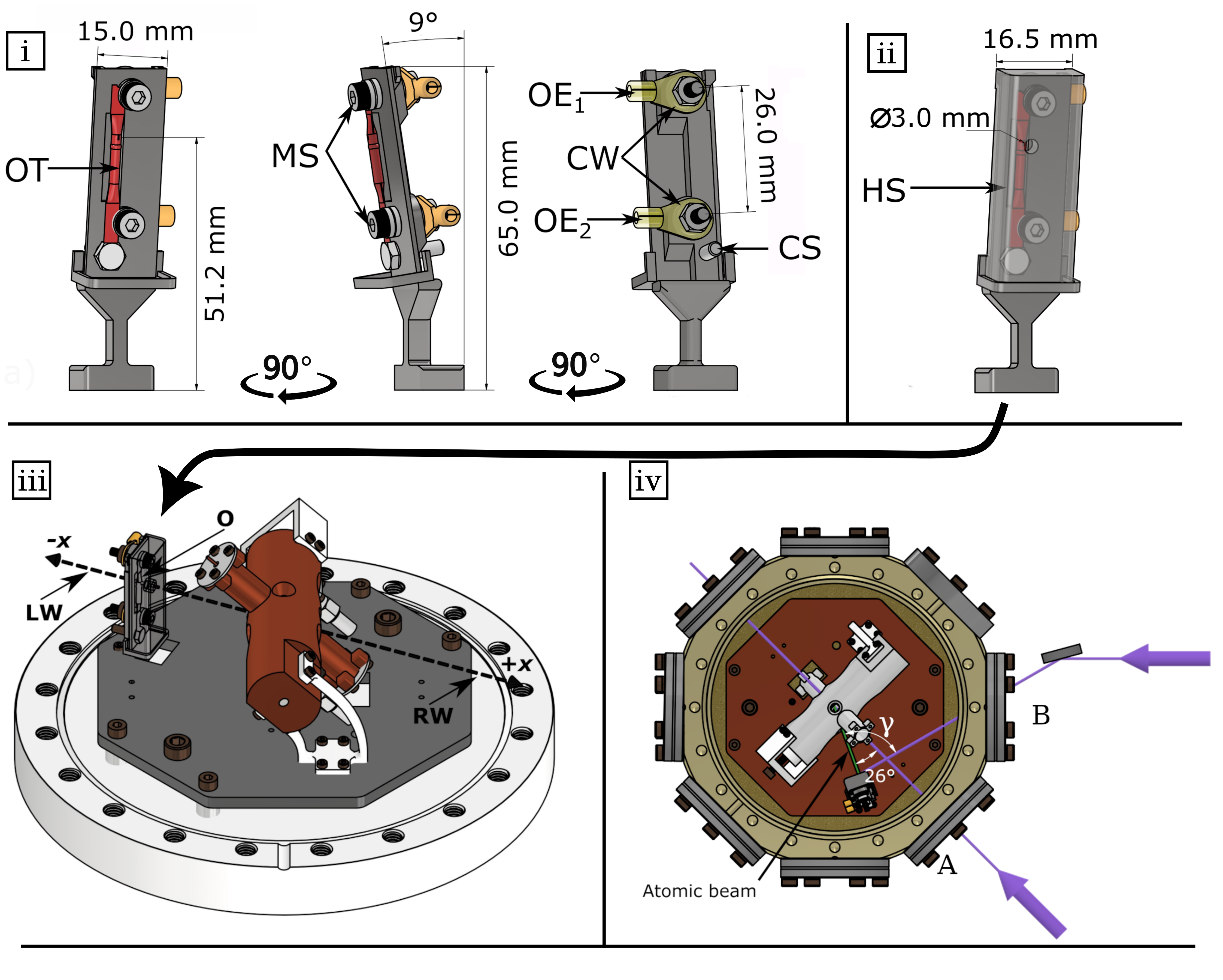}
\vspace{-1.0em} 
\caption{\label{fig:ovendetail}Illustration of the 3D-printed atomic oven assembly and its placement. (i) The view at uncovered oven assembly showing the oven tube (OT) fixed by a pair of metallic screws (MS) and supported by ceramic screws (CS). Copper ring terminals (OE) providing the electrical connection are isolated from the rest by ceramic washers (CW). (ii) Front view with the stainless-steel heat shield (HS) and orifice in place. (iii) Isometric view depicting the installation of the oven and the trap on the bottom DN150CF blank flange of the vacuum chamber. The x-axis of the reference frame used for the FEM simulation extends from the centre of the atomic oven aperture (O) to the centre of the coaxial trap. LW (Left Wall) and RW (Right Wall) denote the inner surfaces of the left and right walls, respectively. (iv) Top view into the vacuum chamber illustrating the orientation of lasers inducing fluorescence of evaporated atoms. The camera for recording the fluorescence is either positioned above the viewport situated on the top flange to face the oven aperture or in front of viewport A.}

\end{figure*}

The compact design of the dispenser, called an oven tube (OT) in the context of the assembly, eliminates the need for large Macor supports\cite{ballance2018short} and allows the use of only a few ceramic parts. This reduces manufacturing costs through simplified fabrication compared to Macor-based assemblies, casting, or conventional machining, and improves scalability. The complete assembly is placed in a chamber consisting of an aluminium chamber and a blank flange as illustrated in Figure~\ref{fig:ovendetail}. On this blank flange, a custom-made baseplate is mounted that holds our oven assembly, a 3D-printed coaxial trap, and an optical cavity.

To minimize thermal leakage to the baseplate, the oven assembly is equipped with a thin support leg. Material reduction on the backside reduces thermal inertia, facilitating fast heating and cooling\cite{kumar2025vacuum}. The OT is both thermally and electrically isolated using standard ceramic split bushes, while metallic screws (MS) passing through these bushes establish electrical connectivity. A heat shield (HS) with a \qty{3}{\milli\metre}-diameter hole (the outer, trap-side, face of which is situated \qty{4.9}{\milli\metre} from the aperture in the OT), encloses the assembly to control atomic beam overspray. For diagnostics, a thermocouple can be spot-welded onto the OT\cite{ballance2018short}. Additionally, the OT support is inclined forward to promote the outflow of molten indium seal, preventing it from coating the internal calcium powder. These modifications are conveniently achieved via 3D-printing.

The 316L SS powder we have chosen for the L-PBF manufacturing method has a grain size of (\qtyrange{15}{45}{\micro\meter}). As supported by previous studies by \citet{simchi2025mastering}, within this size range, the grains exhibit low Hausner ratios ($\sim$ 1.1 to 1.2) and high tapped densities ($>\qty{5}{\gram\per\cubic\centi\meter}$), both indicative of favourable spreading behaviour and dense layer formation. The alloy is UHV-compatible and has been tested at pressures as low as \qty{2.5E-8}{Pa}. 

In the printing process -- the parameters of which are in Table \ref{tab:processing_params} --  a thin layer of 316L powder is spread over the building platform, and a high-power laser selectively melts the powder in an inert argon atmosphere to form each layer of the assembly. Once a layer is completed, the platform is lowered and a new layer of powder is added. This process repeats until the entire atomic oven assembly is built and takes place in an inert gas environment (typically argon) to prevent oxidation of the powder. The final print-out has dimensional tolerances  $\lesssim\qty{100}{\micro\meter}$, which is more than enough for the oven assembly.

\begin{table}[h]
    \centering
    \renewcommand{\arraystretch}{1.2} 
    \begin{tabular}{|l|c|}
        \hline
        \textbf{PARAMETERS} & \textbf{VALUES} \\
        \hline
        Laser power & 200 W \\
        Scan speed & \qty{650}{\milli\meter\per\second}\\
        Layer thickness & \qty{50}{\micro\meter}\\
        Hatch distance & \qty{110}{\micro\meter}\\
        Laser spot size & \qty{75}{\micro\meter}\\
        Scanning strategy & Meander \\
        Pre-heat & Not applied \\
        \hline
    \end{tabular}
    \caption{Key laser processing parameters for 3D printing of oven assembly.}
    \label{tab:processing_params}
\end{table}

\subsection{SEM imaging and EDS spectroscopy of assembly}
\label{sec:SEM imaging and EDS spectroscopy of assembly}

\begin{figure*}
    \centering    
    \includegraphics[width=110 mm]{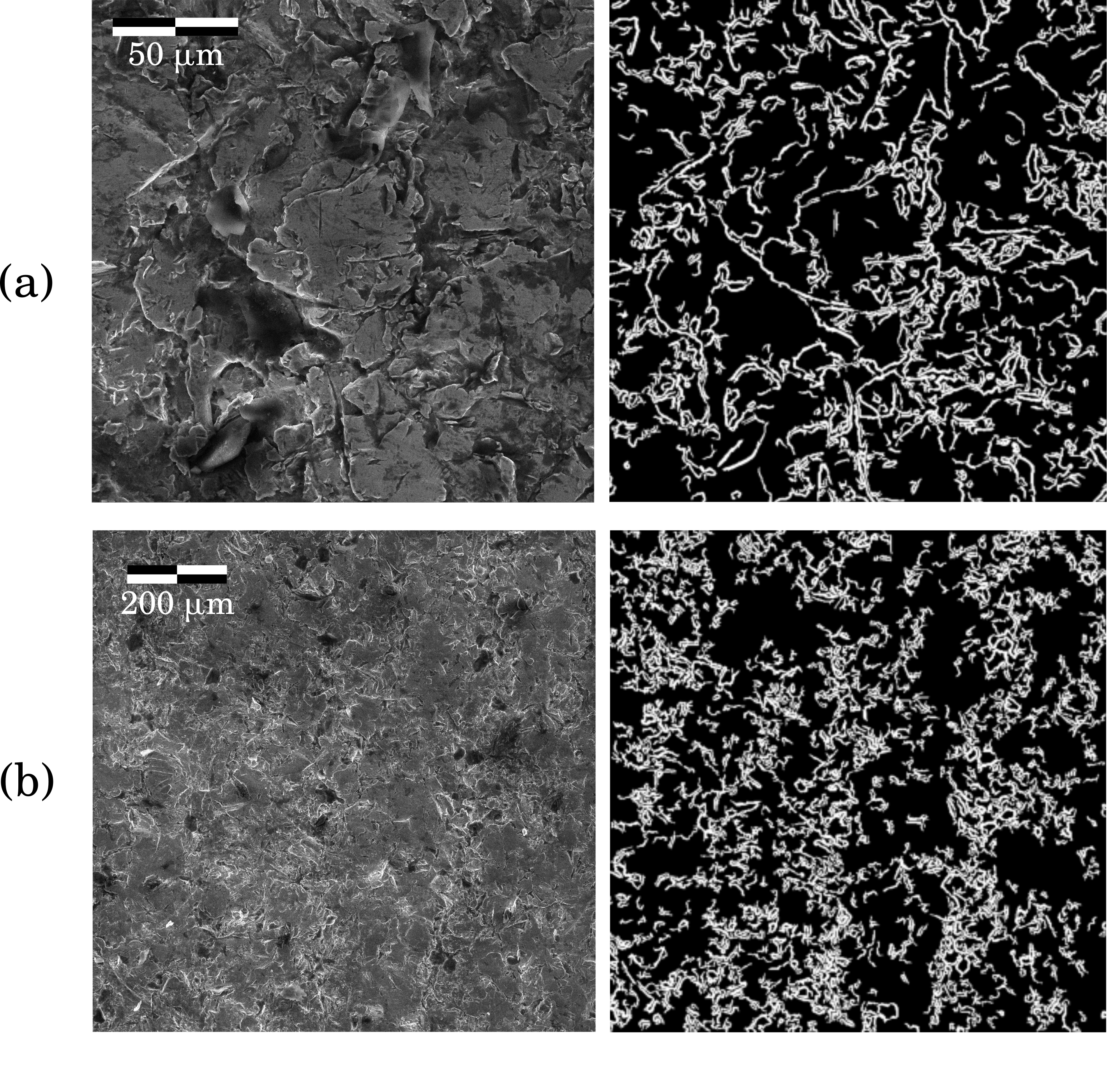}
\vspace{-1.0em} 
\caption{SEM images of surface of print-out and their analyses. (a) and (b) show micrometre and millimetre field views, respectively, coupled with binarized images of cracks. The analyzed pictures had dimensions of $663 \times 666$ pixels.}
\label{fig:SEM Images}
\end{figure*}

Scanning electron microscope (SEM) imaging of the 3D-printed oven assembly reveals important details about surface quality and manufacturing precision. Figure~\ref{fig:SEM Images}a shows a close-up view with a field of 200 \si{\micro\meter}, displaying a well-manufactured, uniform surface at this scale. While minor irregularities are present, they do not significantly impact oven operation.

To understand the surface quality required for the UHV compatibility, we quantified microcracks on the surface that could serve as outgassing sources. Figures \ref{fig:SEM Images}a and \ref{fig:SEM Images}b illustrate the binarization that is the core of our detection workflow. This process begins with contrast-limited adaptive histogram equalization (CLAHE) to enhance local features\cite{reza2004realization}, followed by Gaussian blurring to suppress noise\cite{lindeberg1990scale}. Sobel edge detection then extracts gradient information, and Otsu thresholding binarizes the image\cite{otsu1975threshold}.

The crack density per field of view is $\sim\qty{7E-3}{\per\square\micro\meter}$ for the \qty{200}{\micro\meter} image (Figure~\ref{fig:SEM Images}a) and $\sim\qty{3E-4}{\per\square\micro\meter}$ for the 1 mm image (Figure~\ref{fig:SEM Images}b). Successful maintenance of $<\qty{2.5E-8}{Pa}$ pressure indicates that these microcracks do not compromise UHV integrity.

\begin{figure}
\label{fig3}
    \centering    \includegraphics[width=80mm]{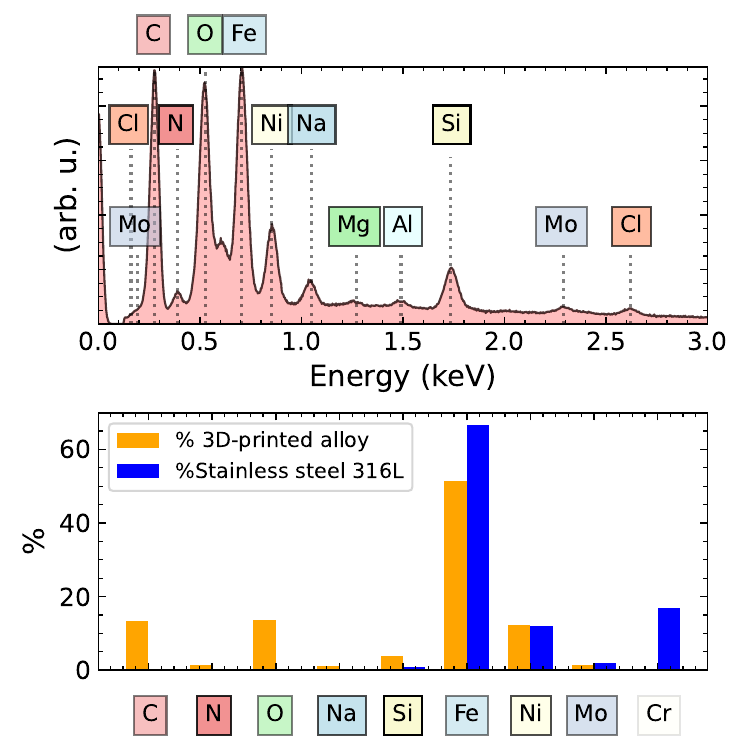}
\caption{\label{fig:3} EDS spectrum of the 3D-printed piece (upper panel), illustrating the elemental composition. The lower panel presents a comparison of the composition percentages between the 3D-printed alloy and stainless steel 316L.}
\end{figure}

The elemental composition of the printed pieces was analyzed using Energy-Dispersive X-ray Spectroscopy (EDS). The measured EDS spectrum, shown in Figure~\ref{fig:3}, shows minor deviations of the 3D-printed alloy from the stainless steel 316L in terms of elemental composition. The excess carbon, nitrogen, and oxygen content and the increase in silicon content result from storage and handling outside the clean-room environment. The significant depletion of chromium content has been reported in other studies to be caused by selective evaporative losses in the process of laser melting\cite{mukherjee2024integrated, murkute2019production}. That may indicate the possibility that miniature gas pockets form during manufacturing. None of the aspects mentioned here eventually appear to have negative effects on the pressure.

At the oven operation temperature of \qty{700}{K}, the vapour pressures of the EDS-identified alloy elements, following \citet{alcock1984vapour}, are extremely low, with  $\sim 2\times 10^{-18}$ Pa (Fe), $\sim2\times10^{-17}$ Pa (Cr), and $\sim 6\times10^{-20}$ Pa (Ni), Mo even lower, corresponding to values 14–17 orders of magnitude lower than that of the evaporant under identical conditions. Therefore, we can safely assume that the calcium is the main element evaporated during the operation of the oven.

The overall assessment indicates that a sufficiently high-quality product has been produced. In particular, the absence of significant defects ensures that the system meets the necessary vacuum and structural requirements. Therefore, the SEM findings affirm that the surface roughness and grain size are well within acceptable tolerances for this particular ion-trap-based application.

\section{Heating dynamics of oven assembly}
\label{sec:Heat simulation of oven assembly}

As mentioned in Section \ref{sec:Introduction}, it is crucial to suppress the heat propagation from the oven to the electron trap. The heat shield shown in Figure~\ref{fig:ovendetail}-ii is designed to minimize heating of surrounding components by blocking thermal radiation. Attached to the oven body, it covers three sides of the tube, exposing less than 8$\%$ of its surface through a small aperture towards the trap centre at $x_\text{c} = \qty{57.8}{mm}$ from the oven as depicted in Figure~\ref{fig:ovendetail}-iii. The origin of the coordinate system is positioned at the oven tube aperture, which is situated $\qty{22.2}{mm}$ from the nearest wall and $\qty{137.8}{mm}$ from the farthest, with a $\qty{18.2}{mm}$ gap from the upper ConFlat (CF) flange.

\begin{figure*}
    \centering
\includegraphics[width=0.9\linewidth]{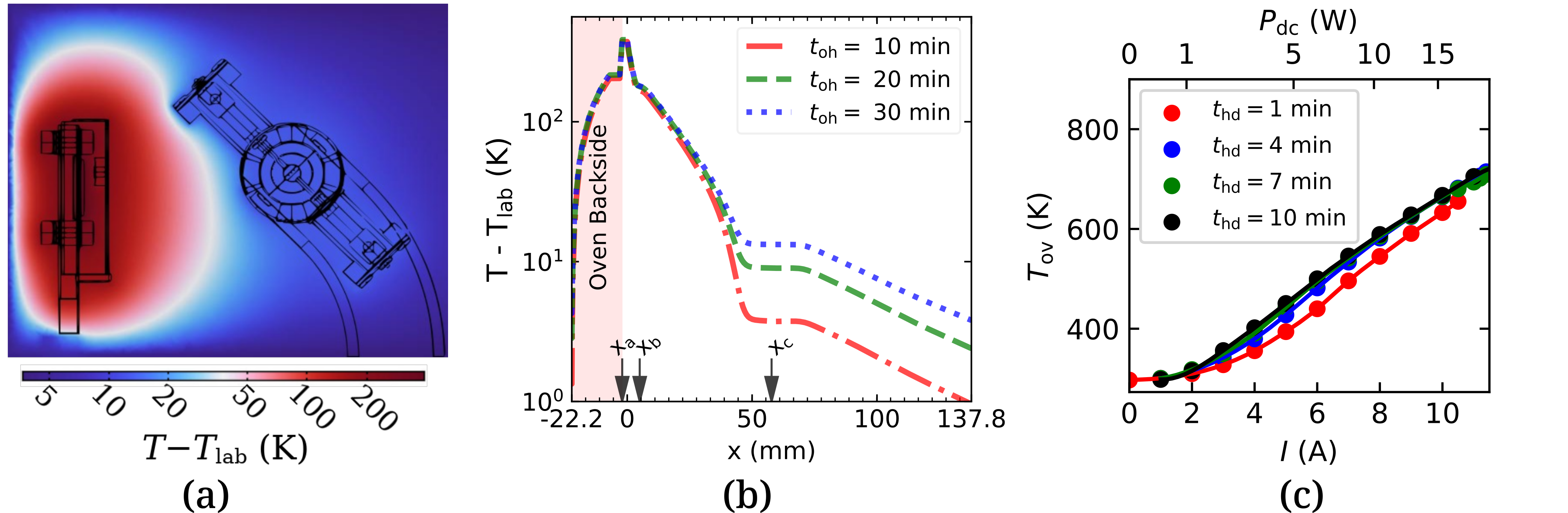}
    \caption{Computational and experimental analysis of thermal behaviour of the oven setup. (a) Computed heat distribution map with heat shield at the heating power of 17.8 W for $t_{\text{oh}}=30$ min. (b) Calculated temperature profile as a function of distance for different heating durations at the same oven power of 17.8 W, referenced to the chamber maintained at the laboratory temperature $T_{\text{lab}} \approx \qty{293}{K}$. Here, $x_\text{a}$ ($x_\text{b}$) denotes the position of the assembly backside (heat shield front). (c) Oven temperature as a function of current for different holding times ($t_\text{hd} = 1, 4, 7, 10 \,\text{min}$). Experimental data points are shown as dots, while smooth spline curves represent fitted trends. The upper axis indicates the corresponding electrical power, mapped directly to the current values. The experiments were performed at the pressures on the order of \qty{E-6}{Pa}.}
    \label{fig:Heat_analysis}
\end{figure*}

Figure~\ref{fig:Heat_analysis}a shows the results of the simulated heat distribution obtained using the Heat Transfer with Surface-to-Surface Radiation multiphysics coupling of Heat Transfer Module in COMSOL Multiphysics 6.1\cite{comsol2024}. This simulation models the thermal behaviour of the atomic source to assess the effects of heat dissipation on nearby components in the vacuum pressure of \qty{E-7}{Pa}. The temperature distribution $T(P_{\text{dc}}; x, t_{\text{oh}})$ was computed along the axis from the oven aperture to the trap centre for varying heating durations $t_{\text{oh}}$ under constant power $P_{\text{dc}}$.

For clarity and reproducibility, the key parameters and modeling choices used in the COMSOL thermal simulations are summarized in Table~\ref{tab:thermal_model}. The table lists the principal physical and numerical model parameters.

\begin{table}[htbp]
\centering
\caption{Thermal model parameters used in COMSOL simulations.}
\label{tab:thermal_model}
\renewcommand{\arraystretch}{1.2}
\begin{tabularx}{\linewidth}{l l X}
\hline
\textbf{CATEGORY} & \textbf{PARAMETERS} & \textbf{VALUES / DESCRIPTIONS} \\
\hline
General conditions 
& Vacuum pressure 
& $10^{-9}$ mbar \\
& Initial temperature 
& 293.15 K (all parts, including chamber walls)\\
& Contact type 
& Perfect thermal contact \\

\hline
Heat capacity $C_p$ 
& Oven and heat shield 
& 475 J\,kg$^{-1}$\,K$^{-1}$ \\
& Chamber and trap 
& Temperature-dependent heat capacity updated during simulation, $C_p \approx \qty{990\pm100}{\joule\per\kilogram\per\kelvin}$, COMSOL default \cite{McBride1993NASA} \\
& Insulator parts 
& 900 J\,kg$^{-1}$\,K$^{-1}$ \\
& Residual gas 
& COMSOL default \\

\hline
Thermal conductivity $k$ 
& Oven and heat shield 
& 44.5 W\,m$^{-1}$\,K$^{-1}$ \\
& Chamber and trap 
& Temperature-dependent thermal conductivity updated during simulation, $k~\approx~\qty{233\pm8}{\watt\per\metre\per\kelvin}$, COMSOL default \cite{Ho1972JPCRD} \\
& Insulator parts 
& 27 W\,m$^{-1}$\,K$^{-1}$ \\
& Residual gas 
& COMSOL default \\

\hline
Surface emissivity 
& Oven and heat shield 
& 0.74 \\
& Chamber and trap 
& 0.095 \\
& Insulator parts 
& 0.42 \\

\hline
Radiation model 
& Radiating surfaces 
& All oven, trap, and inner chamber surfaces facing vacuum \\
& Ambient radiation 
& Outer bottom surface of chamber \\
& Radiation temperature 
& Computed self-consistently \\
& Radiation resolution 
& 256 channels \\

\hline
Meshing 
& Minimum element size 
& 0.975 mm \\
& Maximum element size 
& 22.8 mm \\

\hline
Time stepping 
& Solver 
& BDF (Backward Differentiation Formula) \\
& Time step control 
& Automatic (solver-controlled) \\

\hline
Numerical convergence 
& Iteration error limit 
& 0.1\% in temperature per solver iteration \\

& Mesh refinement study 
& Comparison between original mesh and a mesh with 50\% increased mesh density \\

& Convergence criterion 
& $<0.6$\% difference in temperature rise at the trap center. \\
\hline
\end{tabularx}
\end{table}

The simulations focus on the case where $P_{\text{dc}} = \qty{17.8}{W}$, with heating durations of $t_{\text{oh}} = \qtylist{10;20;30}{\minute}$. In these, the full \qty{17.8}{W} power is used from the start, simplifying the model while capturing essential thermal dynamics. The spatial analysis of the temperature distribution reveals a significantly steeper gradient near the back of the oven (region $\qty{-22.2}{mm} \leq x \leq x_{\text{a}}$) compared to the front side (region $x_{\text{b}} \leq x \leq x_{\text{c}}$), as shown in Figure~\ref{fig:Heat_analysis}a. This indicates strong localized heating near the back of the oven, primarily due to thermal conduction through metallic screws acting as heat bridges.

A significant temperature drop is observed in the region $x_{\text{a}} \leq x \leq x_{\text{b}}$, demonstrating the impact of thermal shielding. In just 5 mm, the temperature falls sharply from approximately 680 to \qty{473}{K} (Figure~\ref{fig:Heat_analysis}b). On the back of the oven tube holder ($x_{\text{a}} = -3$ mm), the temperature reaches about \qty{511}{K}.

In contrast, the forward region ($x_{\text{c}} \leq x \leq \qty{138}{mm}$ in UHV) maintains a stable thermal profile within the range of \qtyrange{297}{306}{K}. Despite being surrounded by electrodes exposed to thermal radiation, heating is minimal even after continuous operation for \qty{30}{\minute}. This shows that thermal shielding effectively limits heat propagation toward the front.

In the real setting, the heating current is supplied through the metallic screws in Figure~\ref{fig:ovendetail}-ii step-wise, while monitoring the oven tube temperature with a thermocouple as described in previous work\cite{kumar2025vacuum}. The evolutions of the oven temperature at various current ramping rates (from \qty{1}{A} per holding time of 10 min to 1 A/1 min) are compared in Figure~\ref{fig:Heat_analysis}c. In the presented setting, the current $I$ (bottom $x$-axis) measured in amperes is mapped to the input power (upper $x$-axis) measured in mW following the empirical formula $P_{\text{dc}}(I) = (0.7 I^4 - 12.2 I^3 + 220.5 I^2 - 101.7 I)$. Introducing a waiting time of several minutes between each current step has been reported to assist in oven outgassing\cite{lechner2010photoionisation} and improves stability in ion-loading procedures\cite{hong2017experimental}. Determining the lower holding time required during baking is therefore important, as it allows the overall heating time of the oven to be reduced to limit the heating of the surrounding, as discussed above in Section \ref{sec:Construction of the oven assembly}. 

Comparative analysis of oven temperature profiles demonstrates that a holding time of $t_\mathrm{hd} = 1$ min is sufficient to achieve nearly the same thermal response as longer durations ($t_\mathrm{hd} = 4, 7, 10$ min). At an input power of $P_\mathrm{dc} = \qty{18}{W}$, the oven temperature was measured at \qty{655}{K}. By raising the power to about \qty{20.5}{W}, the temperature exceeded \qty{683}{K} in under 5 seconds, as reported in the work by \citet{kumar2025vacuum}. This result is promising, since simulations indicate that operating at full power for less than \qty{20}{min} limits the temperature rise to below the favourable threshold of 10 K.

\section{Atomic fluorescence imaging}
\label{sec:Atomic Fluorescence Spectroscopy}

Following detailed characterization of the oven under high vacuum conditions, the fluorescence detection of the generated atomic beam was subsequently evaluated. A 423 nm laser beam was employed to excite the $^{40}\mathrm{Ca}$ atoms from the ground state $4s^{2}\,{}^{1}S_{0}$ to the excited state $4s4p\,{}^{1}P_{1}$ at point O on the aperture hole (as shown in Figure~\ref{fig:ovendetail}-iii) and in the vicinity of the trapping region. Figure~\ref{fig:ovendetail}-iv shows the experimental setup used in this work, including the geometry for directing the atomic beam toward the trap centre and the fluorescence-inspection laser paths at various angles on both the oven side and the trapping region side.  The laser beam is  generated by Sacher Lasertechnik Lion diode laser in Littman/Metcalf configuration, which has the line width on the order of \qty{100}{kHz} in principle. The laser wavelength was measured by HighFinesse WS8-10 wavemeter with sub-MHz resolution. This value is smaller than the natural line width of the fluorescence transition $\sim34~\mathrm{MHz}$ \cite{metcalf1999laser}, so we can state that the apparatus is capable of distinguishing spectral features caused by the Doppler broadening. For laser power measurement, Gentec PH100-Si-HA-OD1-D0 photodiode detector was used. The fluorescence was monitored by an industrial CMOS camera (IDS UI-3360CP-NIR-GL Rev.2).

For proper evaluation, the experiment was initially carried out in two stages. In the first stage, the laser frequency was scanned across the resonance while a constant oven current was maintained to verify that the observed signal comes from fluorescence and not scattering from the surface of the oven. In the second stage, the laser frequency was kept fixed at resonance, and the oven current was varied between 7 A and 11.5 A to study the resulting changes in atomic beam flux and fluorescence intensity. The neutral-atom fluorescence was first inspected at the source side to verify the neutral atom production and then near the trapping-region to validate the experimental setup.

\begin{figure*}
    \centering
    \includegraphics[width=0.9\linewidth]{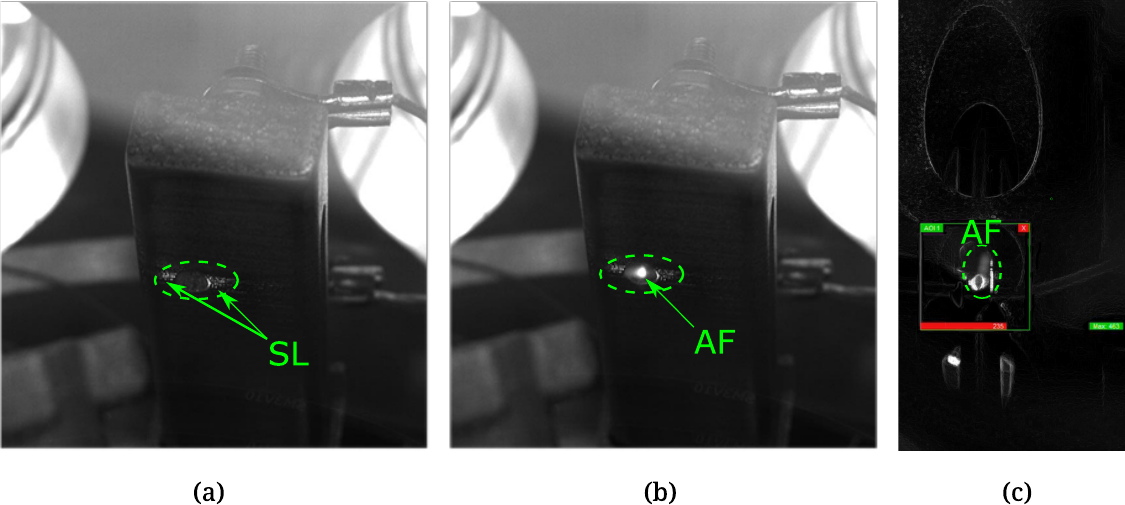}
    
    \caption{Images taken during fluorescence imaging of Ca atom beam. (a) View at the oven aperture through a viewport at top flange. The laser beam partially strikes the heat shield, producing stray light (SL) enclosed within the dashed circular region. (b) The same view as (a) but with atomic fluorescence (AF) spot, clearly confined to the atomic beam region. The spatial distribution and background intensity of the SL remain essentially unchanged between the two images. (c) Atomic fluorescence image near the trap aperture, visible as an O-shaped structure. For orientation, the image was superimposed with one taken at the same angle under ambient lighting.}
    \label{fig:Fluorescence}
\end{figure*}

Figures \ref{fig:Fluorescence}a/b show the oven assembly enclosed by the heat shield during fluorescence imaging through a top viewport (situated on top of the chamber depicted in Figure \ref{fig:ovendetail}-iv). The laser beam is let in the chamber through viewport B so that it hits the heat shield (HS) aperture with diameter of \qty{3}{mm}. The angle $\gamma$ between the laser-beam and the atomic-beam axes is every time around \qty{80}{\degree}. It is observed that, over a laser frequency detuning by several hundred megahertz around the resonance wavelength of \qty{422.7919}{nm}, the fluorescence spot (with the diameter of nearly \qty{1}{mm}) shifts laterally from left to right and back from right to left as the scan direction is reversed. This behaviour is attributed to the Doppler effect, where atoms moving with different velocity components along the laser axis come into resonance at different detunings. 

During this measurement, the laser power was maintained at $\approx\qty{250}{\micro\watt}$. The chamber pressure increased from an initial value of \qty{4.5E-8}{Pa} (before oven baking) to about \qty{8E-7}{Pa} during oven operation. The supply current reached up to 11 A for 3 minutes before being reduced to 8.5 A, at which the fluorescence of the atomic beam remained clearly visible.

In Figure~\ref{fig:Fluorescence}a, we can see the stray light (SL) from the laser beam being scattered from the heat shield surface when the wavelength is out of resonance. The observed pattern does not change when the resonance is established (Figure~\ref{fig:Fluorescence}b) and serves as the evidence that the observed light spot is not the result of laser beam getting misaligned, and hence hitting other surfaces, due to the change of wavelength.



Figure~\ref{fig:Fluorescence}c presents the fluorescence image at the trap centre viewed through viewport A from Figure~\ref{fig:ovendetail}-iv. During this measurement, the chamber pressure increased from \qty{5.1E-8}{Pa} (before oven baking) to approximately \qty{1.1E-6}{Pa}. The oven current was regulated between 10 A and 11.5 A to modulate the fluorescence intensity. In this case, the laser was shot through viewport A as illustrated in Figure~\ref{fig:ovendetail}-iv and its frequency was stabilized at 422.7909 nm with a small residual deviation, which the optimized ARTIQ-based PID regulator maintains within $\pm\qty{2}{MHz}$ with fast settling time; the laser power was maintained at approximately $\qty{270}{\micro\watt}$. Within this range, the measured deviation in the laser beam power remained below $\qty{1}{\micro\watt}$. The observed resonance frequency shift of $\approx1.7$ GHz relative to the value measured at the source side is attributable to the difference in laser beam incident angles (\qty{80}{\degree} vs \qty{26}{\degree} -- see Figure \ref{fig:ovendetail}-iv). The fluorescence intensity recorded at the trap centre (\qty{53}{\milli\metre} from the OT) is three orders of magnitude smaller than that measured close to the HS aperture (\qty{4.9}{\milli\metre} from the OT). In addition to the obvious reduction in atomic flux caused by the beam’s divergence, another contributing factor is that the cone of atomic rays emerging from the oven tube (OT) is partially obstructed by the aperture in the trap resonator body, which can be faintly seen beneath the green rectangle in Figure \ref{fig:Fluorescence}c.

\begin{figure*}
    \centering
    \includegraphics[width=1.0\linewidth]{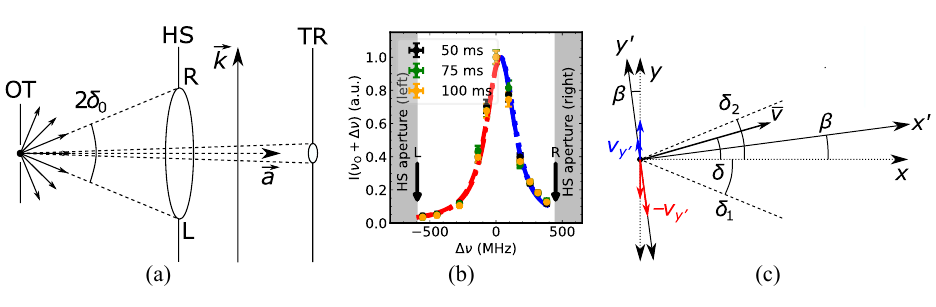}
    
    \caption{Velocity-selective laser-induced fluorescence diagnostics across the HS aperture. (a) CAD-based geometry of the HS aperture, illustrating the acceptance cone directed toward the trapping region (TR) within the atomic emission cone. (b) Normalized spectra fitted with a Voigt profile; shaded regions indicate the velocity selection imposed by the left and right edges of the HS aperture. (c) Schematic of the experimental configuration illustrating asymmetric red- and blue-detuned velocity components relative to the ideal perpendicular configuration, highlighting the measured emission cone angle.}
    \label{fig:Neutral atom fluorescence}
\end{figure*}

Among all atomic beams emerging from the oven, only those that enter the trapping region contribute to ion production. The cone of such rays -- the so-called acceptance cone -- occupies the solid angle of $\sim$\qty{E-4}{sr}, as illustrated in Figure \ref{fig:Neutral atom fluorescence}a. This implies that, from an estimated atomic flux of $\sim10^{12}$ atoms per second emitted from the OT aperture (area $0.2~\mathrm{mm^2}$) at \qty{700}{K}, less than a ten-thousandth of the atomic flux reaches the trap center. This amount is large enough because only tens of ions are expected to be trapped.

The narrow spatial angle of atomic rays entering the trapping region allows us to ignore the angular variation of atomic flux exiting the HS aperture and to concentrate on the study of velocity distribution of rays propagating along the axis $\vec{a}$ of emission cone depicted in Figure \ref{fig:Neutral atom fluorescence}a. Keeping the oven current at 10 A, we repeated the measurement from Figure \ref{fig:Fluorescence}b. As before, we observe that the fluorescence spot resulting from the interaction of the atomic beam with the laser beam $\vec{k}$ shifts from the left (L) to the right edge (R) of the HS aperture as the laser frequency is detuned around the central frequency. The fluorescence spot on the far left is observed at a detuning of $\Delta\nu_\mathrm{L}=\qty{-555\pm12}{MHz}$, while the one on the far right is found at $\Delta\nu_\mathrm{R}=\qty{+382\pm16}{MHz}$. Beyond these boundary values, the fluorescence signal drops abruptly.

To evaluate the spread of velocity components along the laser beam ($\vec{k}$) for atomic rays lying within a cone centred about $\vec{a}$, the fluorescence intensity $I$ at varying frequency detuning $\Delta\nu$ is calculated by integration of the signal measured by a camera within a digital aperture enclosing $\sim\qty{2E-4}{sr}$ region around the spot O in the HS aperture. To avoid intensity distortion caused by saturation of the camera electronics, the images were captured using three different exposure times. The resulting fluorescence spectra are shown in Figure \ref{fig:Neutral atom fluorescence}b. 

In the spectra, the imbalance between $\Delta\nu_\mathrm{L}$ and $\Delta\nu_\mathrm{R}$ is clearly visible. Following the relation $v_{y'}=\Delta\nu \lambda_0$, where $\lambda_0$ is the Doppler-free resonance wavelength\cite{nizamani2010doppler,li2019cascaded,raven2024atoms}, we obtain the boundaries for velocities along the laser beam $v_{y'1} = \qty{-235\pm5}{\metre\per\second}$ and $v_{y'2}=\qty{162\pm7}{\metre\per\second}$. The imbalance between positive and negative bounds is caused by the tilt of the laser beam $\vec{k}$ with respect to the normal to the beam propagation vector $\vec{a}$ by angle $\beta = \qty{90}{\degree}-\gamma=\qty{4.3(0.3)}{\degree}$ as illustrated in Figure \ref{fig:Neutral atom fluorescence}c. In this rotated coordinate frame (X'Y'), where $\vec{k}\,\not\perp\,\vec{a}$, compared to the ideal configuration (XY) with $\vec{k}\,\perp\,\vec{a}$, the red- and blue-detuned boundary velocity components are given by $v_{y'i} = \bar{v}\, \sin(\delta_i+\beta)$, where $\bar{v}$ is the mean speed at 700 K and $\delta_i$ are the propagation angles of the rays on the boundary. Equating this model with the experimentally observed boundaries $v_{y'i}$, we obtain the angles $\delta_1 = \qty{-18.4\pm0.6}{\degree}$ and $\delta_2 = \qty{19.7\pm0.7}{\degree}$. The emission cone aperture is then $|\delta_1| + |\delta_2| = \qty{38.1\pm0.9}{\degree}$.

This cone aperture agrees well with the geometric aperture determined from our CAD model ($2\delta_0$), differing by only \qty{4.1\pm0.9}{\degree}. The mismatch is most probably caused by the difference between the CAD model and the physical realization. The generated atomic beam can be subjected to additional collimation, like in the work by \citet{li2019cascaded}. A simple comparison of the emission-cone and the acceptance-cone apertures illustrates that without such measures, the loading efficiency is suboptimal, despite being sufficient for our research purposes.

\section{Conclusions and outlook}
\label{sec:Conclusions and outlook}

We successfully manufactured a neutral calcium atom beam source (oven) using Laser Powder Bed Fusion technology and validated its performance under experimental conditions. Surface analysis via scanning electron microscopy revealed a crack-related pixel density of approximately \qty{7E-3}{\per\square\micro\meter} over a \qty{200}{\micro\meter} field of view. Based on vacuum pressure measurements reported in our previous work~\cite{kumar2025vacuum}, this level of surface integrity is compatible with ultra-high vacuum operation, despite selective chromium evaporation during the printing process.

When integrated near an electron trap, the oven can operate for approximately \qty{20}{min} at full power (\qty{18}{W}) without exceeding a \qty{10}{K} temperature rise in the trap according to simulations by the finite-element method. In experimental conditions, this is achieved with a controlled heating current ramp of 1~A/min, which is sufficiently slow to enable calcium atom evaporation, as confirmed by our experiments.

Fluorescence imaging confirmed that the neutral calcium atom beam reaches the trapping region. Under two distinct laser incidence angles, we observed both Doppler shifts of the resonance frequency of the $4s^{2}\,{}^{1}S_{0}\!\leftrightarrow\!4s4p\,{}^{1}P_{1}$ transition. Furthermore, detuning-dependent fluorescence imaging of atoms emerging from the oven enabled us to determine the divergence angle of the rays directed toward the electron trap’s trapping region. Although the resulting emission-cone half-angle is approximately \qty{19}{\degree}, the observation of fluorescence within the electron-trap aperture using an industrial CMOS camera, together with the estimation of current of atoms, indicates that a sufficient number of atoms (around \qty{E8}{\per\second}) flows through the trapping region to support effective production of ions and electrons.

Employing 3D printing not only allowed efficient use of space but also delivered a cost-effective and adaptable manufacturing solution. This approach meets the stringent requirements of electron trapping experiments and can be extended to other setups constrained by limited space. Importantly, additive manufacturing imposes no limitations on achieving ultra-high vacuum, establishing it as a viable and competitive alternative to conventional fabrication methods.

\funding{
This work is supported by the Czech Science Foundation (GA\v CR: GA24-10992S), the Charles University Grant Agency (GAUK 295023 and GAUK 131224), and the Czech Ministry of Education, Youth, and Sports (project QM4ST, id. no. EH22\_008/0004572). Additional funding was provided by the Cooperatio Programme of Charles University. IH thanks to the
Technology Agency of the Czech Republic (TAČR: TN02000020) for the support. We also acknowledge previous funding from the University -- the Primus Research Programme (PRIMUS/21/SCI/005). Furthermore, VK acknowledges the European Cooperation in Science and Technology (COST Action: CA17113, Ref.: E-COST-GRANT-CA17113-4b3f9042) for a Short-Term Scientific Mission grant, which facilitated training on a cryogenic linear Paul ion trap at the University of Liverpool.}

\roles{
\textbf{Vineet Kumar}: Conceptualization (equal), Data curation (lead), Formal analysis (lead), Investigation (lead), Visualization (lead), Validation (lead), Writing – original draft; \textbf{Niklas V Lausti}: Software (equal), Validation (supporting), Visualization (supporting); \textbf{Peter K\'{u}\v{s}}: Resources (equal), Writing – review \& editing (supporting); \textbf{Adam Jel\'{i}nek}: Data curation (supporting), Software (equal); \textbf{Ivan Hud\'{a}k}: Formal analysis (supporting); \textbf{David Moty\v{c}ka}: Data curation (supporting); \textbf{Petr Dohnal}: Resources (equal), Writing – review \& editing (supporting); \textbf{Ji\v{r}\'{i} Hajny\v{s}}: Resources (equal), Writing – review \& editing (supporting); \textbf{Radek Pla\v{s}il}: Resources (equal), Writing – review \& editing (supporting); \textbf{Michal Hejduk}: Conceptualization (equal), Funding acquisition (lead), Project administration (lead), Resources (equal), Supervision (lead), Writing – review \& editing (lead)}

\data{
The findings of this study are openly available online\cite{vineetkumarSupportingDataVerification, artiq}. The CAD (STEP) files for the oven assembly and its electronic controller, including the code for laser stabilization and detuning using ARTIQ modules (Zotino and Kasli), are available at the same source\cite{artiq}. A video for a broader audience demonstrating fluorescence in an electron–ion trapping setup, including chamber details, is available\cite{kumar2025eitex}.}

\printbibliography

\end{document}